\documentclass[aps,prb,twocolumn]{revtex4}
\usepackage{graphicx}
\usepackage{dcolumn}
\begin{document}
\title{Coupling between magnetic ordering and structural 
instabilities in perovskite biferroics: A first-principles study}
\author{Nirat Ray and Umesh V. Waghmare}
\affiliation{{Theoretical Sciences Unit}\\
{Jawaharlal Nehru Centre for Advanced Scientific Research}\\
{Jakkur PO, Bangalore 560 064, India}}
\date{\today}
\begin{abstract}
We use first-principles density functional theory-based calculations to investigate
structural instabilities in the high symmetry cubic perovskite structure of rare-earth 
(R $=$ La, Y, Lu) and Bi-based biferroic chromites, focusing on $\Gamma$ 
and $R$ point phonons of states with para-, ferro-, and antiferromagnetic ordering. 
We find that (a) the structure with G-type antiferromagnetic ordering is most stable, 
(b) the most dominant structural instabilities in these oxides are the ones associated 
with rotations of oxygen octahedra, and (c) structural instabilities involving changes 
in Cr-O-Cr bond angle depend sensitively on the changes in magnetic ordering. The dependence of 
structural instabilities on magnetic ordering can be understood in terms of how super-exchange
interactions depend on the Cr-O-Cr bond angles and Cr-O bond lengths.  We demonstrate
how adequate buckling of Cr-O-Cr chains can favour ferromagnetism. Born effective charges (BEC)
calculated using the Berry phase expression are found to be anomalously large for
the A-cations, indicating their chemical relevance to ferroelectric distortions.

\end{abstract}
\maketitle
\section{Introduction}

A ferroic is a material which exhibits spontaneous and switchable ordering of electric polarization
or magnetization or elastic strain. Materials exhibiting more than one of such orderings termed `multiferroics'-
have recently become the focus of much research\cite{Works}. 
Most of the biferroics investigated in recent years are $ABO_3$ oxides with perovskite structure.  
The $d^0-$ness or the zero occupancy of tranistion metal $B$ cation
is known chemically to favor ferroelectricity\cite{Hill1}.
Hence, the availability of transition metal \textit{d}-electrons in the perovskite 
oxides necessary for magnetism, reduces the tendency for off-centering ferroelectric 
distortions\cite{Hill1} making multiferroics relatively rare.
How the ordering of $d-$electronic spins of the $B$ cation influence ferroelectric or other
competing structural instabilities has not yet been explored and understood. The coupling
between magnetic ordering and structural instabilities is expected to involve interesting
physics and is of direct relevance to technological applications\cite{Wood} such as
multiple state memory elements and novel memory media. 

There are at least three families of high temperature biferroic materials.
Bi- based perovskite oxides, like BiMnO$_3$ \cite{Hill3}, BiCrO$_3$\cite{Hill2}, 
and BiFeO$_3$\cite{BFO,Neaton} are known to be promising biferroics.
Ferroelectricity in these materials arises from the stereochemical 
activity of the $6s$ lone pair electrons of Bi. Hexagonal rare earth manganates
LnMnO$_3$ and InMnO$_3$ are biferroics which exhibit improper or geometric
ferroelectricity\cite{VanAken,IMO,Rabe,Mn}. Rare earth chromites LnCrO$_3$ \cite{Serrao,Sahu}
(with Ln = Ho, Er, Tm, Yb, Lu or Y)  have been recently shown to be biferroic;
YCrO$_3$ has been shown\cite{Serrao} to exhibit canted antiferromagnetic behavior below 140 K 
and a ferroelectric transition around 473 K. Similarly, LuCrO$_3$ becomes a canted 
antiferromagnet below 115 K, and is ferroelectric below
488 K \cite{Sahu}. 
The absence of any ferroelctricity in LaCrO$_3$, has been attributed
to the large size of the La$^{3+}$ ion, in comparison with Y$^{3+}$.

However, small values of polarization reported for these 
materials (about 2$\mu$C/cm$^2$ for YCrO$_3$\cite{Serrao} and 6$\mu$C/cm$^2$ of BiFeO$_3$\cite{Neaton})
inspite of large A-cation off-centering distortions remain a puzzle.
More recently\cite{cnr}, a new concept of `local non-centrosymmetry'
in YCrO$_3$ has been proposed to account for the small value of polarization observed. 
Perovskite oxides are known to have many competing structural instabilities\cite{Cockayne}
and this competition is further enriched by the magnetic instabilities.
The coupling and competition between various magnetic and structural ordering can be partly
responsible for weak ferroelectricity or the possibility of local non-centrosymmetry.
Our goal here is to investigate this issue through determination of various structural instabilities
for different magnetic orderings, with a focus on rare earth chromites.
 
We present results of detailed electronic structure and frozen phonon (at the $\Gamma$ and R points)
calculations for a set of five materials 
(LuCrO$_3$, YCrO$_3$, LaCrO$_3$, BiCrO$_3$ and YFeO$_3$)
in the cubic phase with three different magnetic orderings (para-, ferro- and antiferromagnetic).
In Section II, we briefly describe the methods used in calculations here.
In section III, we report results for structural energetics, the electronic density of states (DOS) and 
the Born Effective charges (BEC) for the cubic phase of these materials with different magnetic orderings.
In Section IV, we report results for structural instabilities in chromites and compare them with those in a 
related compound YFeO$_3$. Since the many-electron correlations are important in magnetic oxides,
we estimate their effects on the structural instabilities through use of the Hubbard parameter U \cite{LSDA+U}.
Using our results for structural instabilities, we show how certain nonpolar structural instabilities
can cooperatively stabilize ferromagnetism.
Our work reveals how the structural instabilities of these biferroic oxides depend on the 
size of A-cation (with the same B-cation), and a change in B-cation.
We interpret the results using arguments based on 
superexchange\cite{Anderson}, and the well-known Goodenough-Kanamori rules\cite{Goodenough}.
Finally, we summarize in Section V.
\begin{figure}[htp]
\centering
\includegraphics[scale=0.4]{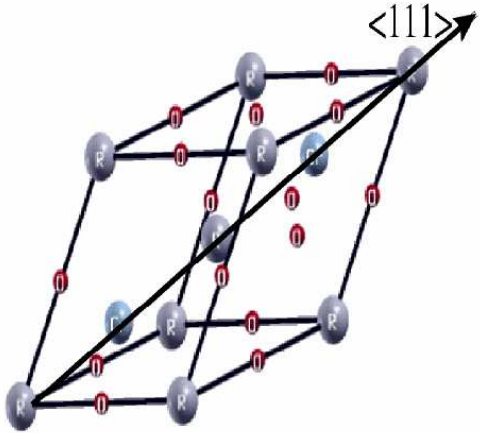}
\caption{Perovskite structure: Cell doubled along the 
$\langle{111}\rangle$ direction, to represent G-type antiferroamgnetic ordering.}
\label{PER}
\end{figure}
\section{Methodology}

Our calculations are based on first-principles pseudopotential based Density Functional Theory 
within a generalized gradient approximation (GGA)\cite{Perdew} as implemented in the PWSCF package\cite{PWSCF}. 
The interaction between ions and electrons is approximated with ultrasoft 
pseudopotentials\cite{Van} treating explicitly 11 electrons [(n-1)s$^2$ (n-1)p$^6$ (n-1)d$^1$ and ns$^2$] 
in the valence shell of Lu(n=6), La(n=6) and Y(n=5). We consider 6 valence electrons for 
Oxygen [2s$^2$2p$^4$] and 14 electrons for Cr [3s$^2$3p$^6$3d$^5$4s$^1$]. We used a plane wave basis 
with kinetic energy cut off of 25 Ryd (150 Ryd) to represent wave functions (density). 
For cubic structures, we sample the Brillouin 
zone using a 5 $\times$ 5 $\times$ 5 Monkhorst Pack Mesh\cite{Monkhorst}, and a denser mesh (6 $\times$ 6 $\times$ 6) 
and higher energy cut-off (30 Ryd) for energy differences.
Phonon frequencies calculated with these larger parameters do not differ much from 
those calculated with a lower cut-off. 
We perform spin polarised calculations by initializing different spins on 
neighbouring magnetic ions; For paramagnetic ordering we initialise a zero spin
on Cr ions. To represent antiferromagnetic ordering, the unit cell is doubled along the 
$\langle{111}\rangle$ direction (see Fig\ref{PER}).
To investigate structural instabilities in the prototype 
cubic structure, we determine its dynamical matrix using frozen phonon calculations. 
We use a finite difference form of the first derivative to compute an element of the 
force constant matrix:
\begin{eqnarray}
K_{i\alpha j\beta}=&&-\frac{\partial F_{i \alpha}}{\partial u_{j \beta}}\nonumber \\ 
                  =&&-\frac{F_{i\alpha}(u_{j\beta}=\Delta) - F_{i\alpha}(u_{j\beta}=-\Delta)}{2\Delta}
\end{eqnarray}
where F$_{i \alpha}$ is the Hellman-Feynman force acting on the \textit{i$^{th}$} atom in $\alpha$ direction, and, 
u$_{j\beta}$ the displacement of the \textit{j$^{th}$} atom in $\beta$ direction with respect to the 
equilibrium structure. We used $\Delta$=0.04 $\AA$, about 1$\%$ of the lattice constant. 
The dynamical matrix is then calculated from the force constant matrix,
\begin{equation}
D_{i \alpha j\beta}=\frac{K_{i \alpha j\beta}}{\sqrt{m_i m_j}},
\end{equation}
whose eigenvalues correspond to the square of the phonon frequencies ($\omega^2$).

In periodic systems, the dynamical charge tensor or Born effective charge tensor can be
defined\cite{Z} as the coefficient of proportionality between
the macroscopic polarization created in direction $\beta$ and a rigid displacement
of the sublattice of atoms {\textit{j}} in direction $\alpha$,
\begin{equation}
Z*_{j,\alpha \beta}= \Omega_o \frac{\partial P^{el}_{\beta}}{\partial u_{j,\alpha}},
\end{equation}
$\Omega_o$ being the unit cell volume.
The polarization is determined using the berry phase formalism \cite{Resta} as implemented 
in the PWSCF package.
\begin{table}
\caption{\label{lattice}Lattice constants of various oxides in the cubic perovskite structure with experimental unit cell volumes.}
\begin{ruledtabular}
\begin{tabular}{cc|ccc}
 &&\multicolumn{3}{c}{Stress(GPa)}\\
\hline
 &a ($\AA$)&PM & FM & AFM \\
 \hline
LuCrO$_3$ & 3.77 & -4.712 & 3.302  & 1.429  \\
YCrO$_3$  & 3.79 & -3.875 & 3.883  & 2.240  \\
LaCrO$_3$ & 3.88 &  -7.043 & -0.083  & -1.564    \\
YFeO$_3$  & 3.83\footnotemark[1]& -15.8  & -5.9 & 2.5 \\
BiCrO$_3$ & 3.85\footnotemark[2]& -15.4  & -9.7 & -11.2\\
\end{tabular}
\end{ruledtabular}
\footnotetext[1]{Ref. \cite{YFeO3}.}
\footnotetext[2]{Ref. \cite{Hill3}.}
\end{table}

\section{Properties of the cubic Perovskite structure}

We have determined electronic structure of RCrO$_3$ compounds (R=Y, Lu, La) in the high 
symmetry cubic structure with different magnetic orderings. This is accomplished through 
calculations with  different initial guesses for atomic
spin polarization and optimizing with respect to spin density.
All our calculations are for the experimental unit cell volumes,
as ferroelectricity is known to be sensitive to lattice constants or pressure (see lattice constants
listed in Table \ref{lattice}).  
In many magnetic compounds, a change of magnetic ordering causes a stress which induces a 
structural distortion\cite{stress}. This concept of `magnetic stress' was introduced
to describe structural phase transitions that are
induced by magnetic ordering, and applied to materials with degenerate (usual e$_g$) orbitals\cite{MnO}.
In these materials as well, a change in magnetic ordering (with fixed lattice parameters)
produces a change in  stress (see Table \ref{lattice}). For LuCrO$_3$ and YCrO$_3$, 
the introduction of spin polarization produces 
a change from compressive stress in the paramagnetic phase to a tensile stress, with stress being
minimum in the antiferromagnetic structure (the lowest energy ordering). 
In LaCrO$_3$, the stress remains compressive with all three magnetic orderings.

\subsection{Electronic structure of YCrO$_3$, LuCrO$_3$ and LaCrO$_3$}

\subsubsection{Cubic Paramagnetic(PM) Structure}
\begin{figure}[htp]
\centering
\includegraphics[width=65mm]{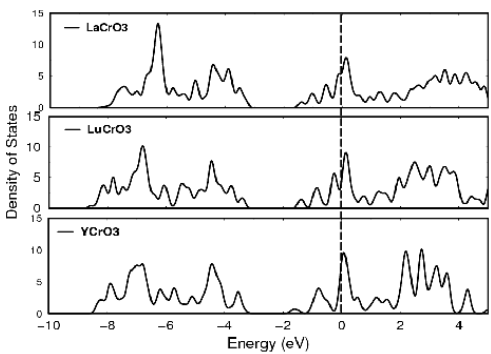}
\caption{Density of States for cubic PM YCrO$_3$, LuCrO$_3$ and LaCrO$_3$. The fermi level (indicated by a dashed line) has been set to zero in all the three cases.}
\label{DOS}
\end{figure}
\begin{figure}[htp]
\centering
\includegraphics[width=75mm]{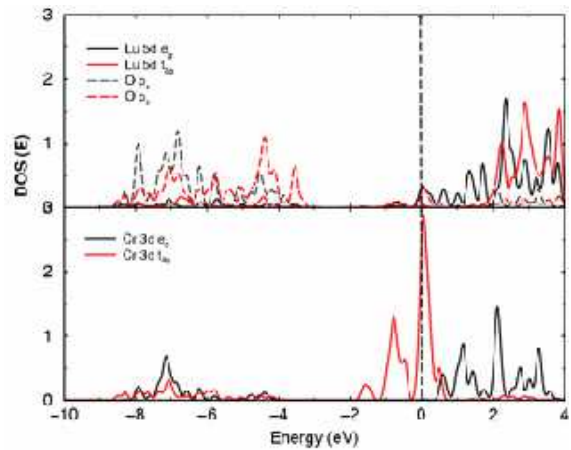}
\caption{Orbital resolved density of states for cubic PM LuCrO$_3$. The high density of states at the Fermi level hints that it is an unstable phase.}
\label{orbresolve}
\end{figure}

First, we present results for the highest symmetry cubic structure with no spin polarization. Although this
state is experimentally inaccessible, it provides a useful reference for understanding the spin-polarized
structures discussed later in the paper. The plotted energy
range is from -10  to 4 eV, and the lower lying semicore
states have been omitted for clarity. In the PM cubic YCrO$_3$, 
LuCrO$_3$ and LaCrO$_3$, (see Fig \ref{DOS}) there is high density of electronic states at the Fermi level,
 driving the system towards a Stoner instability\cite{Stoner}. This suggests that this phase should be 
unstable with respect to spin polarization and/or structural distortions. The contribution of 
various orbitals to the DOS can be understood better by examining the orbital projected
density of states (see Fig \ref{orbresolve}) which show that, the contribution
between -8 to -3 eV is mainly from Oxygen \textit{2p} orbitals. Cr d-orbitals 
contribute predominantly to the peaks at the Fermi level. In contrast, the contribution from the 
Lu d-orbitals is substantial only 2eV above the Fermi level.
\begin{figure}[htp]
\centering
\includegraphics[width=65mm]{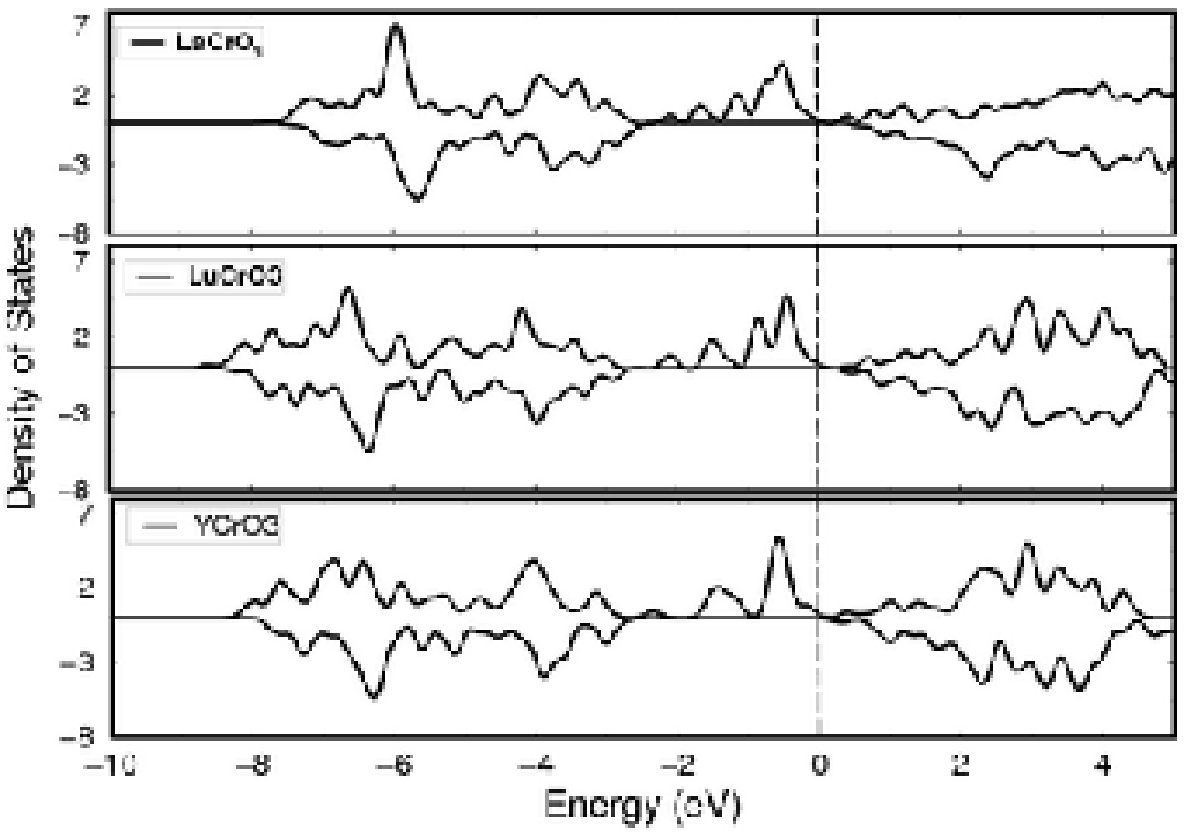}
\includegraphics[width=75mm]{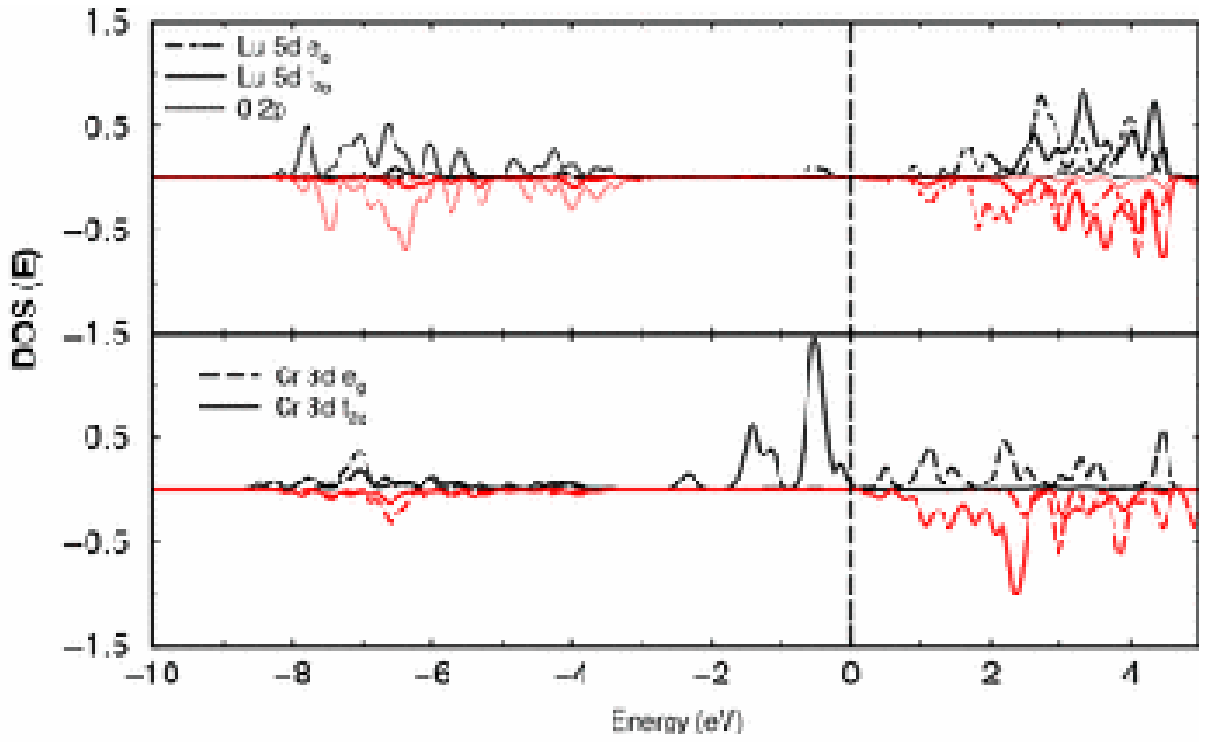}
\caption{Total and orbital resolved Density of states for cubic ferromagnetic YCrO$_3$, LuCrO$_3$ and LaCrO$_3$ with the Fermi level set to zero in all the three cases.}
\label{figure}
\end{figure}
\subsubsection{Cubic Ferromagnetic(FM) Structure}
Ferromagnetic cubic structure is simulated by initializing spins on both Cr ions in the same direction.
In all chromites studied here, the ferromagnetic structures have a magnetic
moment of 3 $\mu_B$ in accordance with the Hund's rule value expected for a d$^3$ configuration.
The majority spins
are represented by the solid line on the positive Y axis, and
the minority spins on the negative Y axis.
The introduction of spin polarization reduces the energy by approximately 2 eV per unit cell. 
The source of stabilization is clear from the density of states (see Fig \ref{figure})
which reveals opening of a gap at the Fermi level.
The states corresponding to non-magnetic atoms are unchanged in comparison
with PM ordering.  The down-spin Cr \textit{3d} states
are split off from the O \textit{2p} states creating a wide gap for the minority states. 
The up-spin Cr \textit{3d} states hybridize with the O \textit{2p} states and there is a very small gap for
the majority carriers. The density of electronic states at the Fermi level is still finite having a small
contribution from the Cr d-orbitals.
This hints that either the ferromagnetic phase may not be the most stable, and that either an
antiferromagnetic (AFM) spin arrangement could lower the energy of the system, or
that the cubic structure is unstable and a structural distortion will lower the energy of the
system. Since Cr$^{3+}$ is a \textit{d$^3$} ion, it is Jahn-Teller inactive, and the structural distortions
(if any) probably involve the A-cation (at the corners), or the oxygen anions.

\subsubsection{Antiferromagnetic Structure}
We simulated antiferromagnetic structure by initializing antiparallel spins on the two Cr ions
in the supercell. 
It is well known that, the superexchange between e$_g$ orbitals of adjacent
ions connected through oxygen with a 180$^o$ metal-oxygen-metal bond angle, is much 
stronger than the interaction between the corresponding t$_{2g}$ orbitals, since the former 
is mediated by stronger dp$\sigma$
bonds as compared to the weaker dp$\pi$ bonds in the latter\cite{Anderson,Goodenough}. So, we expect a superexchange
interaction in which there is a weak coupling between the t$_{2g}$ orbitals of the adjacent Cr
atoms giving rise to an antiferromagnetic interaction. Further, this coupling will be stronger
in the cubic structure as the bond angle between Cr-O-Cr is 180$^0$, as compared to a
distorted structure. 
\begin{figure}[htp]
\centering
\includegraphics[width=70mm]{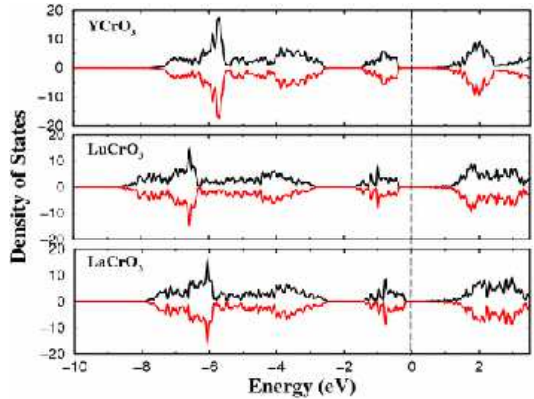}
\caption{Density of states for antiferromagnetic YCrO$_3$, LuCrO$_3$ and LaCrO$_3$.}
\label{AFM}
\end{figure}
From superexchange arguments applied to \textit{d$^3$} configurations, the
structure with G-type antiferromagnetic ordering having rhombohedral symmetry should be most stable.
We consider collinear spins assuming that the canting of the spins would be small.
We find that the AFM structure is lower in energy by about 0.4 eV than the FM phase.
We note that this gain in energy by FM ordering with respect to PM ordering is much more (around 2eV)
than the gain in energy in going from the FM to the AFM structure (see Table \ref{Energies}).
LuCrO$_3$, like YCrO$_3$, is also found to be insulating with the introduction of a gap at
the Fermi level (see Fig. \ref{AFM}). Both spin channels have identical density of states
consistent with an AFM spin arrangement.

\begin{figure}[htp]
\centering
\includegraphics[width=75mm]{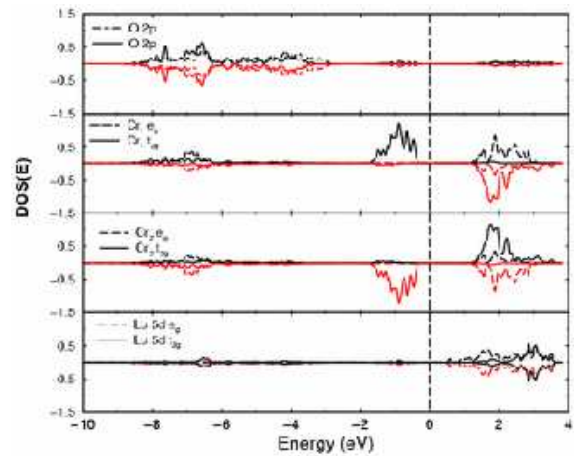}
\caption{Orbital resolved density of states for antiferromagnetic LuCrO$_3$.}
\label{afmpdos}
\end{figure}

From the orbital projected density of states for AFM arrangement, we find that
the t$_{2g}$ orbitals of Cr are fully occupied (and constitute the HOMO), whereas the e$_g$ orbitals are 
unoccupied. Although the LUMO consists of Cr d-orbitals (about 2eV above the Fermi level),
the Lu \textit{d} states also appear within the same energy range (see Fig. \ref{afmpdos}).
\subsection{Born Effective Charge (BEC)}
The effective charge tensors have been calculated from polarization differences between the perfect
and distorted structures in the AFM phase. The anomalous values of
Z* so obtained indicate that a large force is felt by a given ion due to small
macroscopic electric field, thus favoring
a tendency for off-centering and toward a polarized ground state. 
The effective charge of the A-cation (see Table \ref{BEC}) is about the same 
for the three cases. For LaCr$O_3$, we find a larger BEC on
Cr and one of the oxygen atom moving along the bond. This is possibly because of larger
Cr-O bond length (arising from larger size of La cation, see Table \ref{Energies}) 
and correspondingly greater contribution from the long-range charge transfer. 
We expect from this, that the superexchange interaction in LaCrO$_3$ should be stronger as well.

\begin{table}
\caption{\label{BEC}The XX component of Born effective charge tensor
for AFM LaCrO$_3$, LuCrO$_3$ and YCrO$_3$. Nominal charges are indicated in brackets.}
\begin{ruledtabular}
\begin{tabular}{ccccc}
          & Z*$_A$  & Z*$_B$  & Z*$O_x$  & Z*$O_{y,z}$  \\
\hline
LuCrO$_3$ & 4.42(3) & 3.43(3) & -2.56(-2)& -2.62(-2)) \\
YCrO$_3$  & 4.45(3) & 3.44(3) & -2.62(-2)& -2.66(-2)) \\
LaCr$O_3$ & 4.5(3)  & 3.76(3) & -3.82(-2)& -2.22(-2)) \\
\end{tabular}
\end{ruledtabular}
\end{table}
\begin{table}
\caption{\label{Energies}Relative Energies of different Magnetic phases, Cr-O bond lengths, and Neel's temperatures for cubic LuCrO$_3$, YCrO$_3$ and LaCrO$_3$ (Energy of the PM phase has been set to zero).}
\begin{ruledtabular}
\begin{tabular}{cccccc}
            &  PM & FM      & G-AFM   & Cr-O Bond length &         T$_N$ 	    \\
\hline
LuCrO$_3$   & 0.0 & 1.89 eV & 2.3 eV  &    1.88($\AA$)   &    115 K\footnotemark[1] \\
YCrO$_3$    & 0.0 & 2.02 eV & 2.4 eV  &    1.90($\AA$)   &    140 K\footnotemark[1] \\
LaCrO$_3$   & 0.0 & 2.04 eV & 2.5 eV  &    1.94($\AA$)   &    282 K\footnotemark[2] \\
\end{tabular}
\end{ruledtabular}
\footnotetext[1]{Neel's temperature taken from Ref.~\onlinecite{Serrao,Sahu}.}
\footnotetext[2]{Ref.~\onlinecite{Hill3}.}
\end{table}

\section{Structural Instabilities}

\subsection{Coupling with Magnetic Ordering}
In order to represent G-type antiferromagnetic ordering which has been shown to be most favourable
energetically, we use a unit cell doubled along the $\langle{111}\rangle$ direction.
We determine structural instabilities in this structure  with
different magnetic orderings. A single unit cell has  10 atoms which results in 
30 phonon branches: 3 accoustic (which have zero frequency at k=(0,0,0)) 
and 27 optical, some of which are triply degenerate. We are interested mainly in 
optical modes with imaginary phonon frequencies corresponding to instabilities in the structure. 
Doubling the unit cell along the $\langle{111}\rangle$ direction, gives us access to zone boundary
phonon modes (R-point) which form the dominant structural instabilities in this structure,
along with the zone-center modes.

In the paramagnetic phase, both YCrO$_3$ and LuCrO$_3$ exhibit a zone center instability at 116.5 
and 144.8 cm$^{-1}$ respectively, which is a polar mode (with $\Gamma_{15}$ symmetry)
involving mainly the off centering of A-cation. This instability, however, is absent 
in LaCrO$_3$. We find two instabilities in the FM phase: $\Gamma_{15}$ and $\Gamma_{25}$ modes.
The non-polar $\Gamma_{25}$ mode involves oxygen displacements only, and is strongly unstable 
in the FM phase.  The $\Gamma_{15}$ mode involves the A-cation (rare earth ion)
moving in a direction opposite to that of the oxygen cage and Cr atom resulting in a 
ferroelectric polar structural distortion. Note that the Cr atom moves in the same 
direction as the oxygen ion, in contrast to the behavior of Ti ion in BaTiO$_3$\cite{Cockayne} 
and PbTiO$_3$\cite{PTO}, but similar to the behavior of 
Mn in BiMnO$_3$\cite{Mn} and Cr in BiCrO$_3$ \cite{Hill3}.

\begin{figure}[htp]
\begin{center}
\includegraphics[scale=0.5]{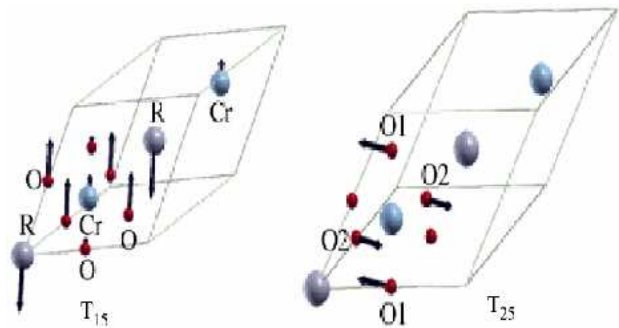}
\end{center}
\caption{Eigenvectors of the unstable $\Gamma$ point phonon modes: $\Gamma_{15}$  and $\Gamma_{25}$ modes}
\label{modes}
\end{figure}

With AFM spin arrangement, for LuCrO$_3$, we find three triply degenerate instabilities, 
at 339.5 $cm^{-1}$, 145 $cm^{-1}$ and the weakest at around 60 $cm^{-1}$. 
For YCrO$_3$, the corresponding instabilities are at 309 $cm^{-1}$, 140 $cm^{-1}$ and 80 $cm^{-1}$ 
respectively. The strongest instability (at around 300 $cm^{-1}$ for the two 
materials) has R$_{25}$ symmetry and corresponds
to rotation of the corner connected oxygen octahedra. The next instability is
the ferroelectric $\Gamma_{15}$ mode (around 140 $cm^{-1}$). The weakest 
instability (60 $cm^{-1}$ for 
LuCrO$_3$ and 80 $cm^{-1}$ for YCrO$_3$) has R$_{15}$ symmetry and involves displacement 
of the A-cations (Lu and Y for our case) and small 
oxygen displacements; these are antiparallel in neighbouring unit cells.
In LaCrO$_3$, we find only two triply degenerate instabilities. The first instability at 
around 220 $cm^{-1}$ corresponding to the oxygen rotations (the R$_{25}$ mode) and the second close
to 18 $cm^{-1}$ having $\Gamma_{15}$ symmetry (see Fig \ref{modes}).

For the rare earth chromites studied in this paper, unstable R-point modes in the AFM phase 
do not change significantly with magnetic ordering. $\Gamma$ point instabilities in contrast depend 
strongly on magnetic ordering. Only the high frequency phonons
are affected in going from para- to antiferromagnetic phase  and all $\Gamma_{15}$ and $R_{15}$ modes
are softer in the FM phase.

The $\Gamma_{25}$ mode which brings about a significant change in the Cr-O-Cr as well as O-Cr-O bond angle
shows a spectacular change with magnetic ordering.
This mode, highly unstable in the ferromagnetic phase (at around 200 cm $^{-1}$) becomes stable 
at around 50 cm$^{-1}$ with para- and antiferromagnetic orderings. This behavior, although 
not as pronounced, is also seen for this mode in LaCrO$_3$. The R$_{25}$ instability, 
also involving a change in the Cr-O-Cr bond angle, is not affected by the change in 
magnetic ordering possibly because the O-Cr-O bond angle still remains unchanged.

Another significant change is observed for the R'$_{25}$ mode,
which involves a movement of the two B-cations (Cr in our case)  in opposite directions.
After introduction of spin-polarization, 
the R$_{25}$' mode (close to 250 cm$^{-1}$ for PM phase) becomes more stable at around 
400 cm$^{-1}$ for the FM and AFM phases.

\begin{figure}[htp]
\centering
\includegraphics[width=80mm]{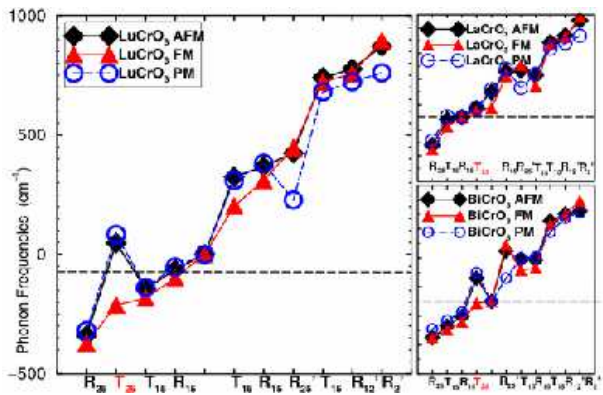}
\caption{Changes in phonon frequencies of cubic LuCrO$_3$ with different magnetic orderings. Insets show similar curves for LaCrO$_3$ and BiCrO$_3$.}
\label{phonon}
\end{figure}

We compare our results with the Bi-based biferroic chromite, BiCrO$_3$ and 
find a similar behavior of the $\Gamma_{25}$ mode here as well. We thus attribute this behavior 
to the B cation (Cr ion for the chromites) and expect it to be their general behavior.
To interpret the general trend in phonon frequencies (see Fig. \ref{phonon}), the following rules apply:

1. The modes which involve a change in the Cr-O-Cr bond angle (as well as O-Cr-O bond angle) are more 
stable in the AFM phases and are relatively less, or unstable, in the FM phase. This behaviour is seen
in the  $\Gamma_{25}$ and  $\Gamma_{15}$ modes for LuCrO$_3$. 

2. Secondly, the modes which involve a change in the Cr-O bond length tend to harden
with the introduction of spin polarization, as observed for the R'$_{25}$ and R'$_{12}$
modes.

\begin{figure}[htp]
\centering
\includegraphics[width=85mm]{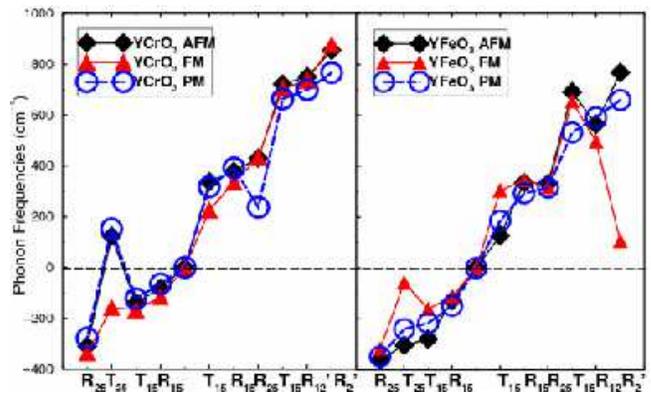}
\caption{Changes in phonon frequencies with changes in B-cation: YCrO$_3$ and YFeO$_3$.}
\label{YFeCr}
\end{figure}

In order to study the effect of change in B-cation (see Fig \ref{YFeCr}), we compare
instabilities in  YFeO$_3$ with YCrO$_3$. We find that a change in magnetic ordering
has an opposite effect on the unstable modes involving a change in Fe-O-Fe bond angle.
A spectacular change is the R'$_2$ mode, an oxygen breathing mode, which softens in the FM phase as
compared to the PM and AFM phases. The  $\Gamma_{25}$ mode also shows a different behaviour, showing a
stabilization with FM ordering. These differences are due to the filled
e$_g$ orbitals in Fe$^{3+}$, which are unoccupied in Cr$^{3+}$, leading to a much stronger superexchange interaction
mediated by the e$_g$ orbitals, and have a different geometry dependence.

\subsection{Stabilization of Ferromagnetic Ordering}
\begin{figure}[htp]
\centering
\includegraphics[width=80mm]{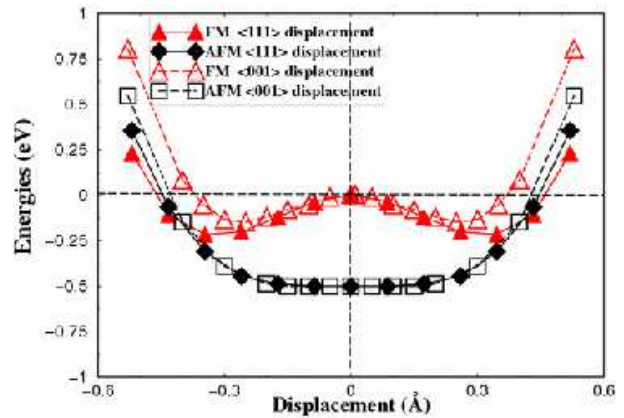}
\caption{Energy vs displacement corresponding to the $\Gamma_{25}$ mode for FM and AFM orderings in LuCrO$_3$.
FM becomes more stable than the AFM state for rhombohedral distortions greater than 0.407 $\AA$.}
\label{Cross_over}
\end{figure}

\begin{figure}[htp]
\centering
\includegraphics[width=65mm]{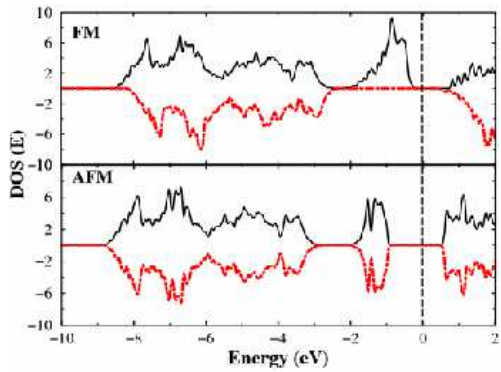}
\caption{Density of states for FM and AFM phases at $\Gamma_{25}$ displacement of  0.407 $\AA$ 
 (see Fig. \ref{Cross_over}, where the FM state is more stable).}
\label{CrossoverDOS}
\end{figure}

The spectacular change in the frequency of the $\Gamma_{25}$ mode with FM ordering, prompts 
us to discuss whether ferromagnetism can be stabilized in these chromites by varying the Cr-O-Cr bond angle. 
Since the $\Gamma_{25}$ mode is unstable only in the FM phase, 
we want to study the effect of freezing in a distortion of this mode.
We have calculated the total energy as a function of $\Gamma_{25}$ displacements in the 
$\langle{111}\rangle$ and $\langle{100}\rangle$ directions, for the FM and AFM phases (see Fig \ref{Cross_over}).
For LuCrO$_3$, we observe a crossover, at a displacement of 0.407 $\AA$, beyond which the FM phase is energetically favoured.
The Cr-O-Cr bond angle, at the crossover point is found to be 153$^{\circ}$, which  
is significantly different from, the value suggested
by Goodenough and Kanamori for ferromagnetic superexchange interaction (130$^{\circ}$) in 1951.
On examining the Density of states beyond the crossover point (see Fig. \ref{CrossoverDOS}),
we find more significant hybridization between Cr and oxygens for the FM phase.
Secondly, the FM phase so stabilized is found to be insulating.

\subsection{Effect of Correlations}

\begin{figure}[htp]
\centering
\includegraphics[width=75mm]{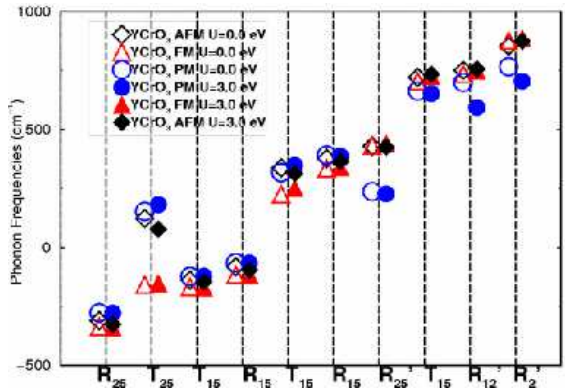}
\caption{Effect of correlations on the phonon modes of YCrO$_3$ with different magnetic orderings. For each mode, data 
on the left of the vertical dashed line represents estimates with GGA and the data on the right represents estimates 
 with GGA+U (U=3.0 eV).}
\label{YCr_plus_U}
\end{figure}

As mentioned earlier in the paper, the LDA+U method has been successfully applied to describe the
electronic structure of sytems containing localized \textit{d} and \textit{f} electrons where
LDA sometimes leads to incorrect results\cite{LSDA+U}, and recently it has been applied to
obtain structural parameters that are in better agreement with experimental results than LDA or GGA.
\cite{lda_plus_u,Neaton}. In this work, we use a value of U = 3.0 eV, adapted from
work on full structural optimization of BiCrO$_3$ \cite{Neaton}.
With the introduction of correlations through U parameter, we find that the modes in FM phase
do not change significantly. Our results for YCrO$_3$ (see Fig \ref{YCr_plus_U})
bear that only $\Gamma_{25}$, R'$_2$ and R'$_{12}$ modes are noticeably affected.
Correlations lead to softening of the R'$_2$ and R'$_{12}$ modes by 50 cm$^{-1}$ , and tend to
harden the $\Gamma_{25}$ mode in the PM phase. In an AFM spin arrangement, these R-point modes are
hardened by approximately 20 cm$^{-1}$ and the $\Gamma_{25}$ instability is softened.

\section{Summary}

In conclusion, we have determined  structural instabilities of
LaCrO$_3$, LuCrO$_3$, YCrO$_3$ and BiCrO$_3$ in their cubic perovskite 
structures with different magnetic orderings. 
Our finding that the G-type antiferromagnetic ordering is most stable can be explained with 
superexchange arguments. Ferroelectric structural instabilities in the 
cubic structures involve A-cation (Lu or Y) displacements, as indicated by the
eigenvectors of the ferroelectric $\Gamma_{15}$ modes and an anomalous BEC of the A-cation.
We find that certain phonon frequencies depend sensitively on magnetic ordering:
the modes involving a change in bond-angle are stable (harder) with the antiferro- and 
paramagnetic ordering than in the FM state; on the other hand, the modes involving a change in Cr-O bond 
length are softer in the paramagnetic phase and comparable in the FM and AFM states. 
The $\Gamma_{25}$ oxygen mode brings about a significant change in the Cr-O-Cr bond angle and
is highly unstable in the FM phase, and corresponding structural distortion leads to stabilization 
of ferromagnetic ordering in these chromites. Among the competing structural instabilities
the antiferrodistortive instability ($R_{25}$ mode) is the strongest. 
Electron correlations are found to have little effect on the unstable phonon modes, but 
result in a slight change in a few of the stable phonon modes in the PM phase.
We note that the effects of magnetic ordering on structural instabilities are quite different 
(in fact, opposite sometimes) in YFeO$_3$ with respect to chromites, and 
a more detailed study is required to understand such couplings in ferrites.
Origin of small polarization and/or local non-centrosymmetry\cite{cnr} is probably from
the relatively weak ferroelectric instabilities and their competition with various structural 
magnetic instabilities, and our work should be useful in formulating a phenomenological analysis
of the same.

\section{Acknowledgments}
Nirat Ray thanks JNCASR for Summer Research fellowship Programme and Joydeep Bhattacharjee for discussions.
UVW is thankful to Professor C N R Rao for stimulating discussions and encouragement for this work and
acknowledges use of central computing facility and financial support from the Centre for 
Computational Materials Science at JNCASR.


\end{document}